\documentclass[lettersize,journal]{IEEEtran}
\usepackage{amsmath,amsfonts}
\usepackage{algorithmic}
\usepackage{algorithm}
\usepackage{array}
\usepackage{textcomp}
\usepackage{stfloats}
\usepackage{url}
\usepackage{verbatim}
\usepackage{graphicx}
\usepackage{orcidlink}
\usepackage{cite}
\usepackage{multirow}
\usepackage{makecell}
\usepackage[caption=false,font=normalsize,labelfont=sf,textfont=sf]{subfig}
\usepackage{algorithm}
\usepackage{multicol}
\begin{document}

\title{Novel Optimized Designs of Modulo 2\textsuperscript{n} + 1 Adder for Quantum Computing}

\author{Bhaskar Gaur \orcidlink{0000-0001-6738-6890}, Himanshu Thapliyal \orcidlink{0000-0001-9157-4517},~\IEEEmembership{Member,~IEEE}
\thanks{Manuscript received April 19, 2021; revised August 16, 2021. This research used resources of the Oak Ridge Leadership Computing Facility, which is a DOE Office of Science User Facility supported under Contract DE-AC05-00OR22725. (Corresponding Author: Himanshu Thapliyal)

The authors are with the University of Tennessee, Knoxville, TN 37996 USA (e-mail: hthapliyal@utk.edu) 
}}



\maketitle

\begin{abstract}
Quantum modular adders are one of the most fundamental yet versatile quantum computation operations. They help implement functions of higher complexity, such as subtraction and multiplication, which are used in applications such as quantum cryptanalysis, quantum image processing, and securing communication. To the best of our knowledge, there is no existing design of quantum modulo (2\textsuperscript{n}+1) adder. In this work, we propose four quantum adders targeted specifically for modulo (2\textsuperscript{n}+1) addition. These adders can provide both regular and modulo (2\textsuperscript{n}+1) sum concurrently, enhancing their application in residue number system based arithmetic. Our first design, QMA1, is a novel quantum modulo (2\textsuperscript{n}+1) adder. The second proposed adder, QMA2, optimizes the utilization of quantum gates within the QMA1, resulting in 37.5\% reduced CNOT gate count, 46.15\% reduced CNOT depth, and 26.5\% decrease in both Toffoli gates and depth. We propose a third adder QMA3 that uses zero resets, a dynamic circuits based feature that reuses qubits, leading to 25\% savings in qubit count. Our fourth design, QMA4, demonstrates the benefit of incorporating additional zero resets to achieve a purer \text{$|0$}⟩ state, reducing quantum state preparation errors. Notably, we conducted experiments using 5-qubit configurations of the proposed modulo (2\textsuperscript{n}+1) adders on the IBM Washington, a 127-qubit quantum computer based on the Eagle R1 architecture, to demonstrate a 28.8\% reduction in QMA1's error of which: (i) 18.63\% error reduction happens due to gate/depth reduction in QMA2, and (ii) 2.53\% drop in error due to qubit reduction in QMA3, and (iii) 7.64\% error decreased due to application of additional zero resets in QMA4. 
\end{abstract}

\begin{IEEEkeywords}
Quantum circuit, quantum computing, quantum adder, quantum modulo adder, NISQ, FTQ
\end{IEEEkeywords}

\section{Introduction}
Quantum computing offers significant advantages over classical computing in the fields of optimization, scientific computation, machine learning, and security. To fully harness these quantum applications, developing efficient quantum computation circuits is crucial. In particular, quantum modulo arithmetic circuits are necessary building blocks for some of the quantum applications such as quantum image processing, quantum cryptanalysis, and pseudo random number generation\cite{jiang2014quantum, ceschini2023modular, roetteler2017quantum, putranto2022another}. 

Quantum modulo arithmetic circuits are specialized circuits specifically supporting modulo operations. These operations ensure that the result stays in (0, modulus-1) range, maintaining closure under operations such as addition, subtraction, and multiplication. In these circuits, the quantum modulo adder is the most useful, enabling the derivation of other quantum modulo arithmetic circuits such as multiplication and exponentiation. With the advent of quantum computing, several designs of quantum full adders and quantum modulo adders have been proposed \cite{cuccaro2004new, draper2004logarithmic, rodney2005, thapliyal2013design}. However, to the best of our knowledge, there is no existing design targeted and optimized in particular for quantum modulo (2\textsuperscript{n} + 1) adder. Developing a quantum modulo (2\textsuperscript{n} + 1) adder can augment existing quantum modulo arithmetic circuits with quantum modulo (2\textsuperscript{n} + 1) circuits for subtraction, multiplication, and exponentiation.

The Quantum Modulo (2\textsuperscript{n} + 1) Adder (QMA) finds applications in pseudo random number generation, Residue Number Systems (RNS), and quantum cryptanalysis \cite{ceschini2023modular, mohan2016residue, roetteler2017quantum, putranto2022another}. However, creating an efficient algorithm for (a + b) modulo (2\textsuperscript{n} + 1) addition has been a challenging problem. One popular approach is the (a + b + 1) modulo (2\textsuperscript{n} + 1) adder, which provides the closest alternative to the original problem statement. This approach also offers flexibility in correcting the output by decrementing one of the input values. 

Our contributions are as follows: We propose, to the best of our knowledge, the first Quantum Modulo (2\textsuperscript{n} + 1) Adder QMA1. Additionally, we have optimized QMA1 to reduce the number of quantum gates in QMA2, which benefits both Noisy Intermediate Scale Quantum (NISQ) and Fault Tolerant Quantum (FTQ) machines. QMA1 and QMA2 are based on static quantum gates, hence satisfying reversibility. Our proposed design also provides the regular Sum along with Modulo (2\textsuperscript{n} + 1) Sum, increasing its potential usefulness. To further reduce qubit count, we propose two dynamic quantum modulo (2\textsuperscript{n} + 1) adders that use non-reversible zero reset operations. The proposed adder QMA3 utilizes zero resets to reduce ancillae count. In the proposed adder QMA4, we introduce zero resets twice to combat errors introduced due to state preparation. We experiment by implementing the four proposed designs of quantum modulo (2\textsuperscript{n} + 1) adder on IBM Washington, a 127-qubit Eagle R1 architecture-based quantum computer. We demonstrate a gradual reduction of 28.8\% in error, establishing the effectiveness of our systematic optimization methodology.

\IEEEpubidadjcol

\section{Background}
\label{Background}
\subsection{Quantum Gates}
\begin{figure}[!t]
	\centering
	\subfloat[]{\includegraphics[width=0.12\textwidth]{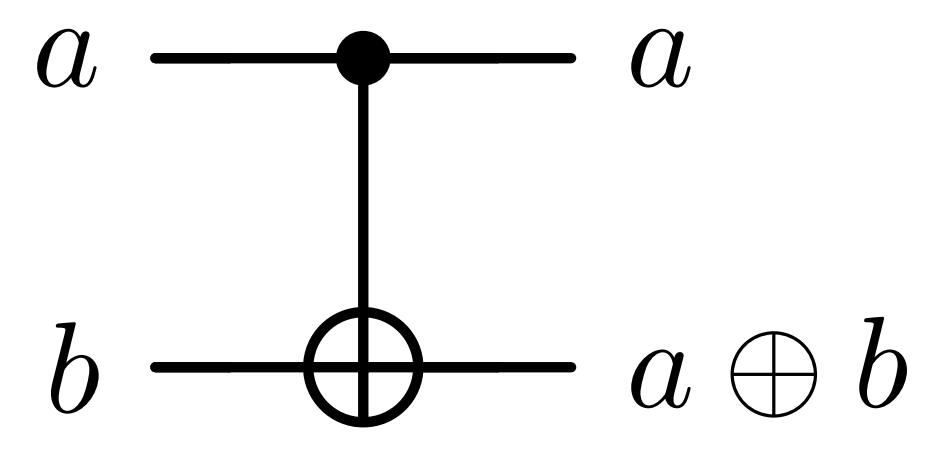}%
		}
	\hfil
	\subfloat[]{\includegraphics[width=0.12\textwidth]{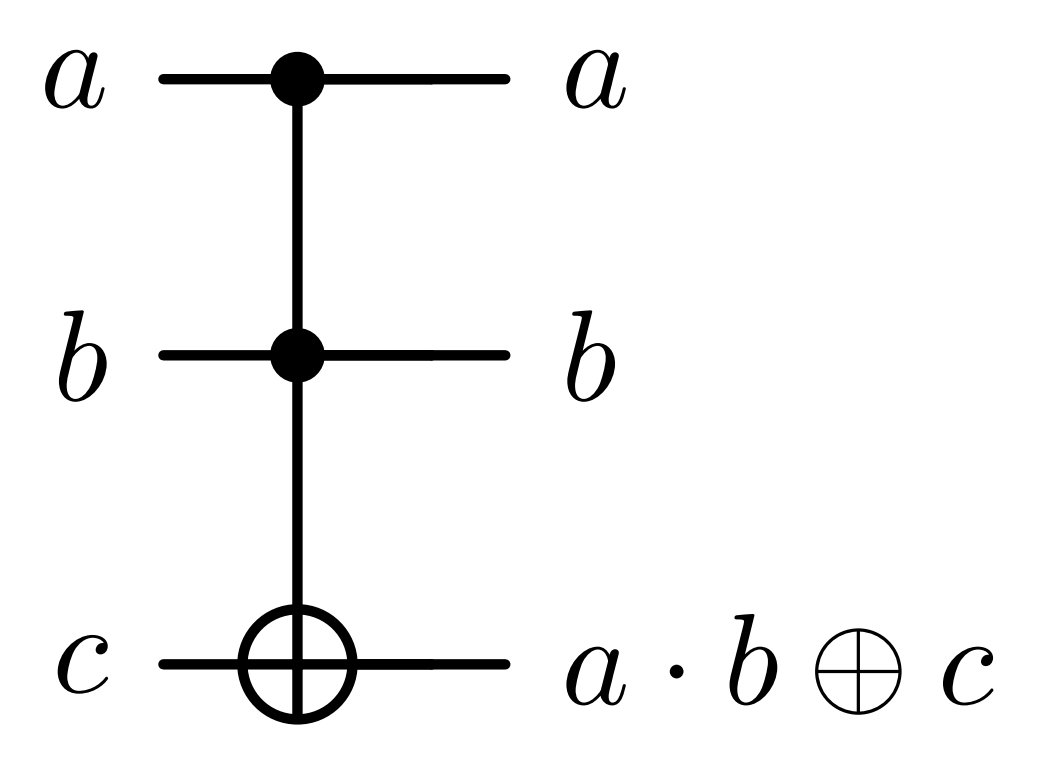}%
		}
	\caption{Quantum gates: (a) CNOT Gate. (b) Toffoli Gate.}
	\vspace{-0.3cm}
	\label{fig:qgates}
\end{figure}

Quantum gates are the fundamental building blocks that perform reversible unitary transformations on atleast one qubit \cite{nielsen2002quantum}. Sequences of quantum gates form quantum circuits. Quantum circuits sometimes need ancillary qubits (or ancillae) to provide constant inputs, which can later be processed to hold intermediate computation steps or the final output. We now discuss the two quantum gates utilized in creating the proposed designs. The first gate is the CNOT or Feynman gate, represented in Figure~\ref{fig:qgates}(a). It is a two input gate that transforms the initial state \text{$|$}A, B⟩ into \text{$|$}A, A$\oplus$B⟩, effectively enabling XOR logical operation. The second gate is the Toffoli gate, also known as the CCNOT gate (double controlled NOT). As depicted in Figure~\ref{fig:qgates}(b), the Toffoli gate converts three input qubits \text{$|$}A,B,C⟩ to three output states \text{$|$}A, B, A$\cdot$B$\oplus$C⟩, thereby executing AND/NAND logical operations, depending on \text{$|$}C⟩. As the Toffoli gate is more resource intensive than CNOT, we measure its gate count and depth separately. Here, depth is defined as the number of gate layers in the circuit that can be executed sequentially.

\vspace{-0.15cm}
\subsection{Modulo (2\textsuperscript{n} + 1) addition}
The modulo (2\textsuperscript{n} + 1) addition operation takes two inputs, a and b, and generates Modulo Sum such that $0\leq{a,b,Modulo Sum}<{2\textsuperscript{n} + 1}$. As shown in Equation \ref{equation:ModAddition}, if the Modulo Sum equals or exceeds (2\textsuperscript{n} + 1), it yields (a + b) mod (2\textsuperscript{n} + 1), focusing solely on the remainder after performing division with the modulus. It has the potential to handle Fermat's prime numbers, implement quantum cryptanalysis, and facilitate other modulo (2\textsuperscript{n} + 1) operations such as subtraction, exponentiation, and multiplication \cite{roetteler2017quantum, putranto2022another}. The utilization of the modulus (2\textsuperscript{n} + 1) can lead to computational efficiency when used in conjunction with other moduli like 2\textsuperscript{n} and (2\textsuperscript{n} - 1) within a residue number system (RNS) \cite{mohan2016residue}.

\begingroup
\setlength\abovedisplayskip{0pt}
\begin{equation} \label{equation:ModAddition}
	\begin{aligned} 				  
		&Modulo Sum=\begin{cases}
			a + b, & \text{if}\;{(a+b)}<{2\textsuperscript{n} + 1} \\
			0, & \text{if}\;{(a+b)}={2\textsuperscript{n} + 1} \\
			(a + b)mod(2\textsuperscript{n} + 1) & \text{if}\;{(a+b)}>{2\textsuperscript{n} + 1}
		\end{cases} \\[10pt]
		&\text{where }\;0\leq{a,b,Modulo Sum}<{2\textsuperscript{n} + 1} \\
	\end{aligned}
\end{equation}
\endgroup

However, creating an efficient algorithm for modulo (2\textsuperscript{n} + 1) addition is a challenging problem. We pursue another approach by optimizing (a + b + 1) modulo (2\textsuperscript{n} + 1) adder instead, which is identical to the original problem statement if one of the input values is decremented by one.

\subsection{Error Metrics}
\label{Error Metrics}
We explain the metrics employed for quantifying the extent of error introduced by modulo (2\textsuperscript{n}+1) adders when executed on quantum computers. The most fundamental metric is the Error Distance (ED), defined in Equation \ref{equation:ED} as the discrepancy between the ideal Sum ($S_{ideal}$) and the real Sum ($S_{real}$). We then use Equation \ref{equation:MED} to calculate the Mean Error Distance (MED) by taking the mean of ED observed across all input combinations of \text{$|0$}⟩ and \text{$|1$}⟩ basis states (N). Finally, we normalize MED by the maximum possible output of the circuit ($S_{max}$), as shown in Equation \ref{equation:NMED}, to derive NMED, a near size-independent metric \cite{liang2012new}.
\begin{equation} \label{equation:ED}
	ED = \left | S_{ideal} - S_{real} \right |
\end{equation}
\vspace{-1.3cm}
\begin{multicols}{2}
	\begin{equation} \label{equation:MED}
		MED = \frac{\sum ED}{N}
	\end{equation}\break
	\begin{equation} \label{equation:NMED}
		NMED = \frac{MED}{S_{max}}
	\end{equation}
\end{multicols}

\section{Proposed Designs}
In order to achieve modulo (2\textsuperscript{n} + 1) addition, we utilize Equation \ref{eq:1} to construct the quantum modulo (2\textsuperscript{n} + 1) adder QMA1. The two inputs, A = a\textsubscript{n} a\textsubscript{n-1} ... a\textsubscript{0} and B = b\textsubscript{n} b\textsubscript{n-1} ... b\textsubscript{0}, are of size (n+1) qubits, while their Sum is an (n+2) qubit integer Sum = A + B = S\textsubscript{n+1} S\textsubscript{n} ... S\textsubscript{0}. Equation \ref{eq:1} achieves optimized modulo (2\textsuperscript{n} + 1) addition using different components of the Sum.

\vspace{-0.5cm}
\begin{multline} \label{eq:1}
	(a + b + 1) mod (2^n + 1) = (a + b) mod\, 2^n + S\textsubscript{n+1} 2^n +\\ \overline{(S\textsubscript{n+1} \vee S\textsubscript{n})}\;\;\;
	\text{if}\;0\leq{a,b} < {2^n+1}
\end{multline}

\subsection{Static Quantum Modulo (2\textsuperscript{n} + 1) Adders}
In this section, we propose two Quantum Modulo (2\textsuperscript{n} + 1) Adders (QMA) based on static quantum gates that satisfy reversibility.
\subsubsection{QMA1: Proposed Design}
\begin{figure*}[h]
	\centering
	\includegraphics[scale=0.34]{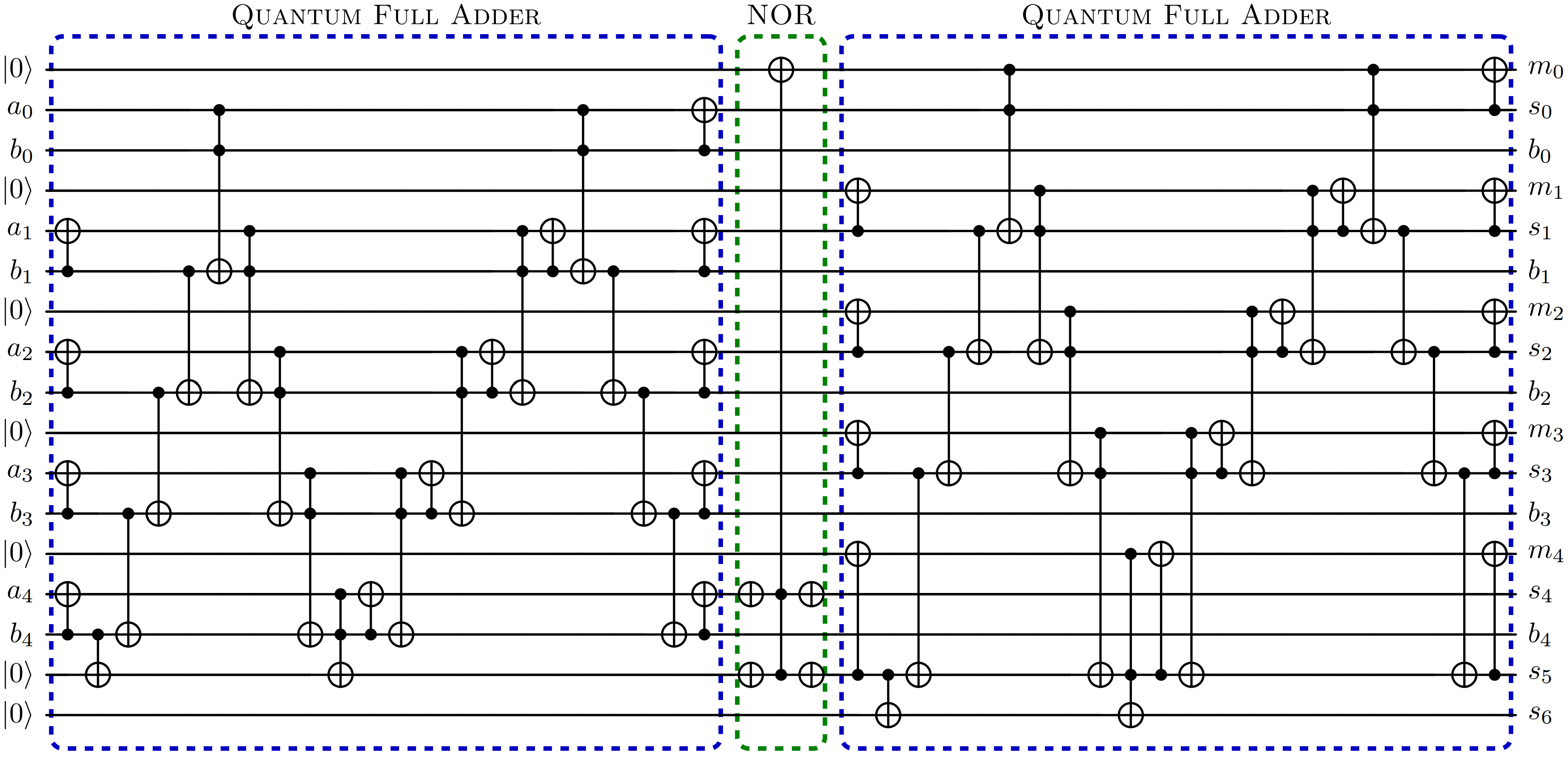}
	\vspace*{-2mm}
	\caption{Proposed quantum modulo (2\textsuperscript{n} + 1) adder QMA1 for n=4 in a 5-qubit configuration. The QMA1 takes the inputs (a\textsubscript{0}:a\textsubscript{4}) and (b\textsubscript{0}:b\textsubscript{4}) to generate Modulo Sum (m\textsubscript{0}:m\textsubscript{4}). The Sum (s\textsubscript{0}:s\textsubscript{6})  and input (b\textsubscript{0}:b\textsubscript{4}) are retained. QMA1 is the basic interpretation of Algorithm \ref{alg:1}, using quantum full adder to calculate (a + b) modulo 2\textsuperscript{n} as intermediate Sum, followed by NOR operation to compute $\overline{(S\textsubscript{n+1} \vee S\textsubscript{n})}$. Another quantum full adder adds these with S\textsubscript{n+1} 2\textsuperscript{n}.}
	\label{fig:QMA1}
\end{figure*}

To implement the Equation \ref{eq:1} in quantum circuits, we use the Algorithm \ref{alg:1}. The algorithm utilizes a quantum full adder to generate the (n+2)-bit Sum = S\textsubscript{n+1} S\textsubscript{n} ... S\textsubscript{0}. We create an intermediate (n+1)-bit Sum$'$ = S\textsubscript{n+1} S\textsubscript{n-1} ... S\textsubscript{0} by substituting the component S\textsubscript{n} with S\textsubscript{n+1}. The Sum$'$ is equivalent to the addition of (a + b) modulo 2\textsuperscript{n} and S\textsubscript{n+1} 2\textsuperscript{n}. To create the NOR component $\overline{(S\textsubscript{n+1} \vee S\textsubscript{n})}$ in the proposed quantum modulo (2\textsuperscript{n} + 1) adder, we utilize a quantum NOR gate constructed using a Toffoli gate and four NOT gates between the two full adder stages of QMA1 in Figure \ref{fig:QMA1}. 

\begin{algorithm}
	\caption{: Calculation of $(a + b + 1) mod (2^n + 1)$}\label{alg:1}
	\begin{algorithmic}
		\REQUIRE $n \geq 1$
		\ENSURE $0\leq{a,b} < {2^n+1}$
		\STATE 1. $Sum = a + b = S\textsubscript{n+1} S\textsubscript{n} ... S\textsubscript{0}$
		\STATE 2. {\textsc{Omit S\textsubscript{n} from Sum to generate:}}\\ $\;\;\;\; Sum' = S\textsubscript{n+1} S\textsubscript{n-1} ... S\textsubscript{0}$
		\STATE 3. $Modulo Sum = Sum' + \overline{(S\textsubscript{n+1} \vee S\textsubscript{n})}$
	\end{algorithmic}
	\label{1}
\end{algorithm}

The resulting NOR component is then placed at the least significant bit (LSB) of a (n+1) qubit input initialized with \text{$|0$}⟩. This (n+1) qubit input is added to the Sum$'$ using another quantum full adder in Figure \ref{fig:QMA1}. As the output or Modulo Sum M cannot exceed the modulus, its size is limited to (n+1)-bits and is represented as M = M\textsubscript{n} ... M\textsubscript{0}. We use the carryless quantum full adder designed by Thapliyal et al. to save quantum gates, and its double-V structure increases noise resilience by reducing the gap between qubit operations, resulting in lesser qubit idling\cite{thapliyal2013design, bgaur2023noise}. 

\vspace{-0.4cm}

\subsubsection{QMA2: Optimized Proposed Design}
\label{Optimized}
In this Section, we propose another quantum modulo (2\textsuperscript{n} + 1) adder QMA2 to optimize the QMA1 by replacing the quantum full adder with a quantum half adder in the second stage of QMA1. This optimization is possible because the size of the input $\overline{(S\textsubscript{n+1} \vee S\textsubscript{n})}$ being added to Sum$'$ is only one bit, which generates a Carry upon addition serving as second input for the remaining bits of Sum$'$. By adopting this optimization, QMA2 significantly reduces quantum gates and eliminates the need for an extra qubit for MSB in Sum for reversibility.

\vspace{-0.4cm}
\begin{figure}[h]
	\centering
	\includegraphics[width=1\linewidth]{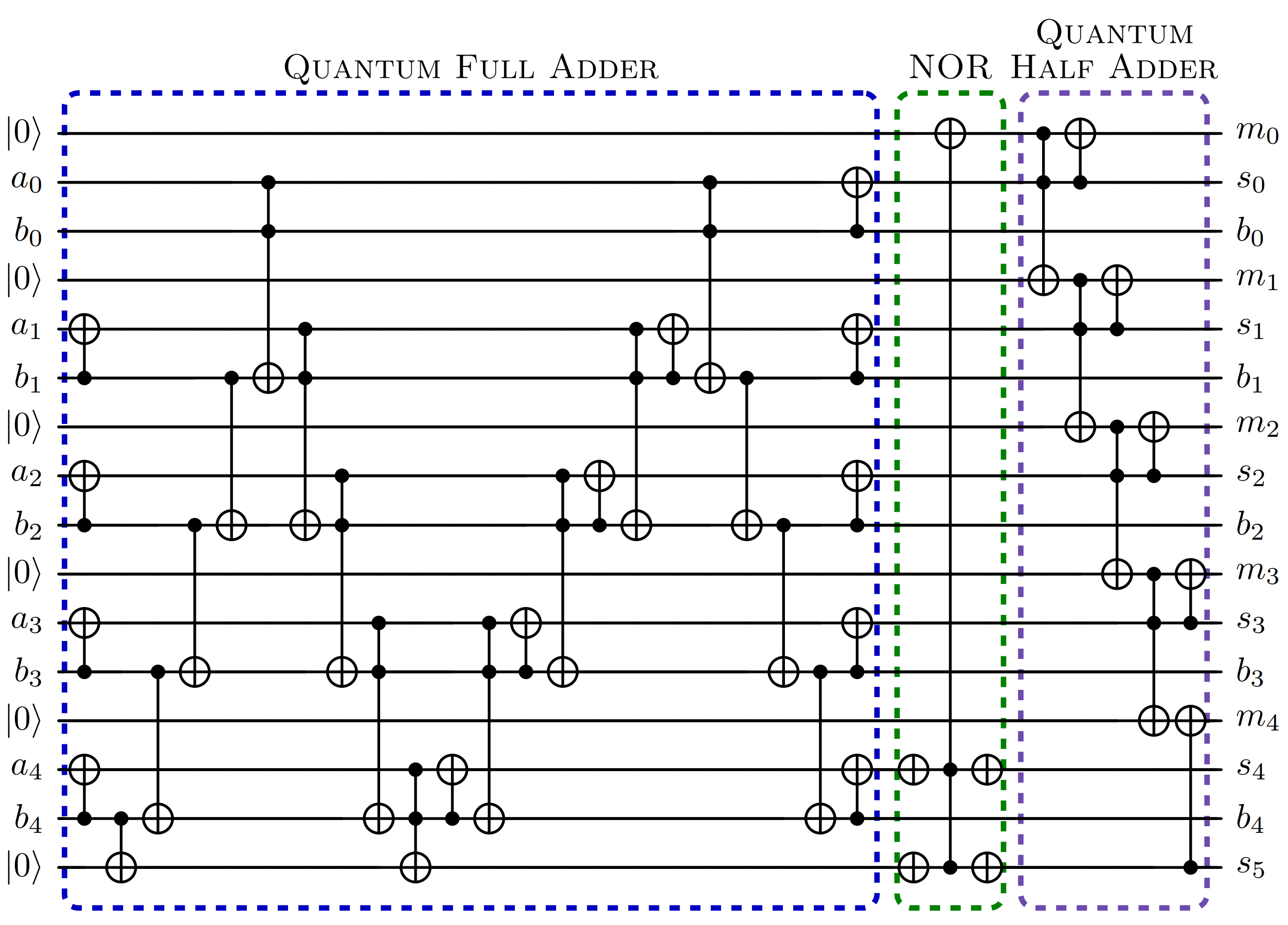}
	\caption{Optimized quantum modulo (2\textsuperscript{n} + 1) adder QMA2 for n=4 in a 5-qubit configuration. The inputs (a\textsubscript{0}:a\textsubscript{4}) and (b\textsubscript{0}:b\textsubscript{4}) are used to generate Modulo Sum (m\textsubscript{0}:m\textsubscript{4}) and Sum (s\textsubscript{0}:s\textsubscript{5}). Unlike QMA1, QMA2 uses a quantum half adder instead of a full adder to reduce gate count.}
	\label{fig:QMA2}
\end{figure}
\vspace{-0.5cm}

\subsection{Dynamic Quantum Modulo (2\textsuperscript{n} + 1) Adders}
In this section, we propose two quantum modulo (2\textsuperscript{n} + 1) adders optimized using zero resets, a dynamic circuit based non-reversible feature.
\subsubsection{QMA3: Zero Reset based Error Reduction}
\label{DynamicCkt}
\begin{figure*}[h]
	\centering
	\includegraphics[width=0.90\linewidth,height=0.4\linewidth]{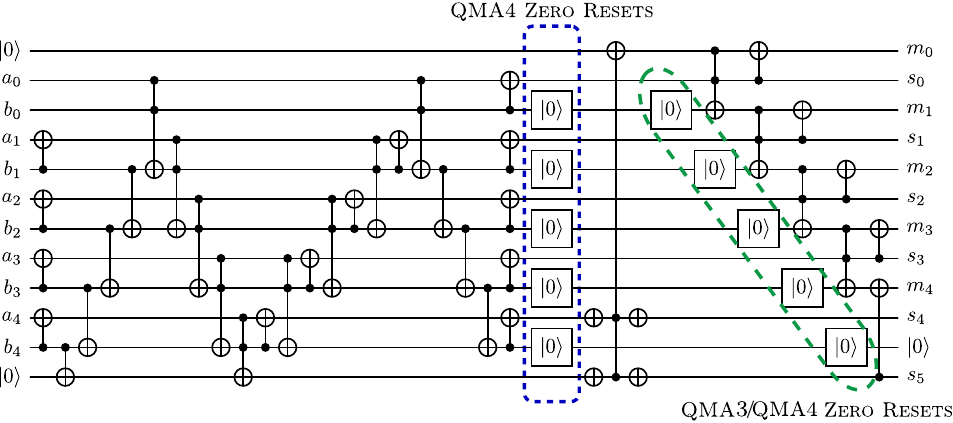}
	\vspace*{-3mm}
	\caption{Proposed quantum modulo (2\textsuperscript{n} + 1) adders QMA3 and QMA4 for n=4 in a 5-qubit configuration. Zero reset helps in reusing a qubit by resetting it to \text{$|0$}⟩. The zero resets in the box (a) are in both QMA3 and QMA4, while zero resets in the box (b) are only in QMA4. QMA3 uses the zero resets in the box (a) to help reuse input (b\textsubscript{0}:b\textsubscript{4}) to generate Modulo Sum (m\textsubscript{0}:m\textsubscript{4}). Sum (s\textsubscript{0}:s\textsubscript{5}) is generated in parallel as before. QMA4 utilizes zero resets in box (b), in addition to zero resets present in box (a) for QMA3, to reduce quantum state preparation errors.}
	\vspace*{-0.4cm}
	\label{fig:QMA34}
\end{figure*}

To further optimize the qubit utilization in our quantum modulo adder, QMA2, we utilize dynamic circuit based zero reset feature. Specifically, we target the qubits carrying the reverse computed input \text{$|b$}⟩, which become redundant once they exit the full adder responsible for calculating the intermediate (n+1)-bit Sum$'$ in QMA2. By resetting these qubits to \text{$|0$}⟩, we reuse them to execute half addition alongside the NOR component $\overline{(S\textsubscript{n+1} \vee S\textsubscript{n})}$. This innovative design is shown in Figure \ref{fig:QMA34} and referred to as QMA3. The introduction of zero resets reduces the qubit count by (n-1) and yields two distinct advantages, although at the cost of reversibility.

Firstly, the qubits carrying input \text{$|b$}⟩ between the control and target qubits of the Toffoli gates within the half adder are now reset to zero for use as target qubits. This elimination of input \text{$|b$}⟩ helps reduce design constraints during physical mapping of the qubits on a quantum computer. Secondly, the zero-initialized ancillary qubits, employed later in the QMA2 by the half adder, are susceptible to decoherence errors caused by idling \cite{bgaur2023noise}. However, QMA3 resolves this issue effectively by resetting the qubits to \text{$|0$}⟩ closer to the half adder, ensuring more accurate \text{$|0$}⟩ states. Both these innovations significantly reduce qubit usage while maintaining the performance of our quantum modulo (2\textsuperscript{n} + 1) QMA3 adder.

\subsubsection{QMA4: Reducing Quantum State Preparation Errors}
\label{StatePrep}
The ripple carry mechanism within the quantum half adder of QMA3 is still susceptible to errors originating from inaccurate \text{$|0$}⟩ states. These inaccuracies can propagate from the least significant bit (LSB) to the most significant bit (MSB), potentially compromising the entire computation. Correcting quantum state preparation errors to achieve purer states requires significant quantum resources in the form of extra quantum gates, measurements, and ancillae qubits. However, we implement a more straightforward approach that minimizes design changes by implementing multiple reset operations to achieve purer \text{$|0$}⟩ state and reduce state preparation errors. 
\vspace{-0.1cm}
\begin{equation} \label{equation:StatePrepError}
	\rho = (1 - \delta)^k |0\rangle\langle0| + \delta^k |1\rangle\langle1|
\end{equation}
\vspace{-0.1cm}
Considering 'k' consecutive zero resets, each with a bitflip error probability of $\delta$, we achieve the best possible mixed state density matrix ($\rho$), as shown in Equation \ref{equation:StatePrepError}. Depending on $\delta$, the output state can range from pure to fully mixed for 0 $\leq \delta < \frac{1}{2}$. In QMA3, we introduce additional zero resets, effectively approximating the behavior described in Equation \ref{equation:StatePrepError} for k=2. These extra resets are shown in the box (b) of Figure \ref{fig:QMA34}, supplementing those already in the box (a) of Figure \ref{fig:QMA34} for QMA3. This design is called the proposed quantum modulo (2\textsuperscript{n} + 1) adder QMA4. Like QMA3, QMA4 is also non-reversible but achieves a more accurate MSB result. By reducing the likelihood of errors arising from inaccurate \text{$|0$}⟩ states within the half adder, QMA4 demonstrates greater error resilience than QMA3, as elaborated in the results section.
\vspace{-0.2cm}

\section{Resource Usage Comparison}
Table \ref{table:comparison} presents a detailed comparison of the quantum resource usage and depth of the four proposed quantum modulo (2\textsuperscript{n} + 1) adders. For n = 4, QMA2 exhibits a significant reduction of 37.5\% in the CNOT count compared to QMA1, with values of 25 and 40, respectively. Additionally, the CNOT depth is reduced by 46.15\% in QMA2, dropping from 26 in QMA1 to 14. Similarly, the Toffoli count and depth show a reduction of 26.3\% from 19 in QMA1 to 14 in QMA2. Hence, QMA2 represents a significant improvement over QMA1 regarding quantum resource conservation and depth reduction. QMA3, on the other hand, reduces the qubit requirement by 25\% from 16 to 12 compared to QMA2, while the CNOT and Toffoli usage remains the same. QMA3 also uses 5 zero resets, which double to 10 in QMA4.

\begin{figure}[h]
	\centering
	\includegraphics[scale=0.34]{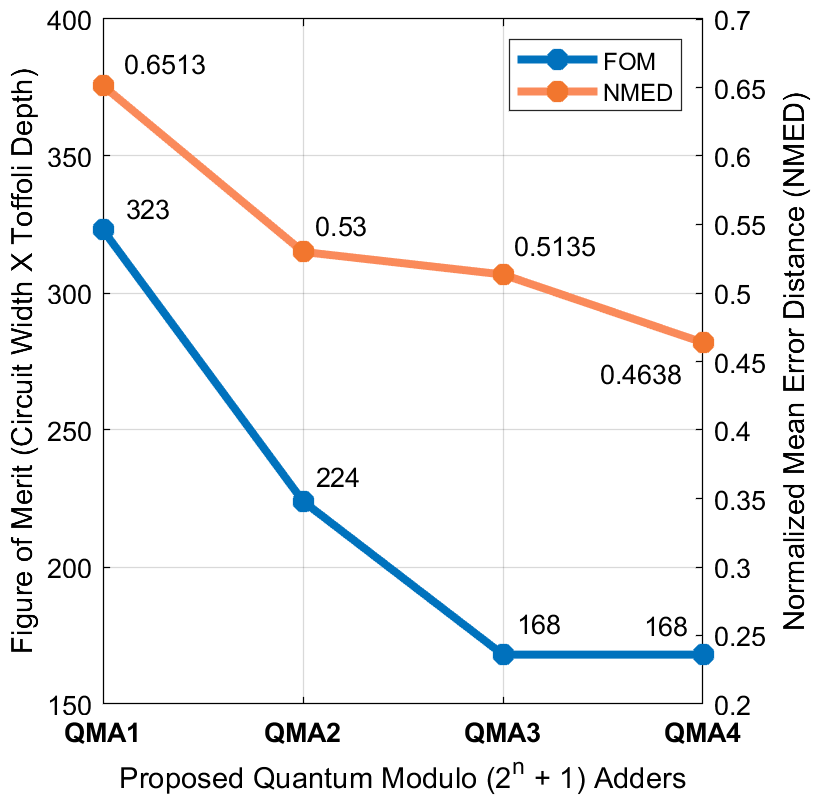}
	\vspace*{-2mm}
	\caption{The above graph compares the proposed quantum modulo (2\textsuperscript{n} + 1) adders, QMA1 to QMA4, for n=4 in a 5-qubit configuration, with NMED representing error and Figure Of Merit (FOM) representing resource usage. FOM = (Circuit Width x Toffoli Depth). For both NMED and FOM, lower is better. From QMA1 to QMA3, the NMED falls along with the reduction in FOM, showing the impact of lower quantum resource usage. However, the 7.64\% reduction in NMED from QMA3 to QMA4, without any drop in FOM, reflects reduction in quantum state preparation errors due to additional resets.}
	\label{fig:Graph}
\end{figure}
\vspace{-0.1cm}

\begin{table*}[h]
	\centering
	\caption{Comparison of Proposed Quantum Modulo (2\textsuperscript{n} + 1) Adders for Quantum Resource Usage and Reduction in NMED.}
	\label{table:comparison}
	{\normalsize%
		\begin{tabular}{|l|l|c|c|c|c|c|c|c|c|} 
			\hline
			\textbf{Category} & \textbf{Adder} & \textbf{Qubits} & \textbf{Zero} & \multicolumn{2}{c|}{\textbf{Depth}} & \multicolumn{2}{c|}{\textbf{Gate Count}} & \textbf{NMED} & \textbf{NMED Drop \%} \\ 
			\cline{5-8}
			&  &  & \textbf{Resets} & \textbf{CNOT} & \textbf{Toffoli} & \textbf{CNOT} & \textbf{Toffoli} &  & \textbf{w.r.t. QMA1} \\ 
			\hline
			\multirow{2}{*}{\textbf{Static }} & \textbf{QMA1} & 3n+5 & 0 & 6n+2 & 4n+3 & 10n & 4n+3 & 0.6513 & 0.0 \\ 
			\cline{2-10}
			& \textbf{QMA2} & 3n+4 & 0 & 3n+2 & 3n+2 & 6n+1 & 3n+2 & 0.53 & 18.63 \\ 
			\hline
			\multirow{2}{*}{\textbf{Dynamic }} & \textbf{QMA3} & 2n+4 & n+1 & 3n+2 & 3n+2 & 6n+1 & 3n+2 & 0.5135 & 21.16 \\ 
			\cline{2-10}
			& \textbf{QMA4} & 2n+4 & 2n+2 & 3n+2 & 3n+2 & 6n+1 & 3n+2 & 0.4638 & 28.8 \\
			\hline
		\end{tabular}
	}%
	\vspace{-0.2cm}
\end{table*}

\section{Experimental Results}
We conduct experiments using IBM Washington, a 127-qubit quantum computer based on Eagle R1 architecture, on the proposed quantum modulo (2\textsuperscript{n} + 1) adders for n=4 in a 5-qubit configuration. We create individual circuits initialized with all possible input combinations of \text{$|0$}⟩ and \text{$|1$}⟩ basis states in IBM Qiskit for each proposed adder \cite{Qiskit}. To maintain uniformity and ensure that the circuits encounter equivalent noise and environmental conditions, we execute them consecutively on the quantum computer with a frequency of thousand shots each, using Qiskit's default qubit placement strategy. Upon execution, we identify the output with the highest frequency and compare it against the accurate or ideal Modulo Sum, calculated as (a + b) modulo (2\textsuperscript{n} + 1). We then compute the Normalized Mean Error Distance (NMED), as explained in Section \ref{Error Metrics}, a crucial metric whose decline helps us track the error reduction in the proposed adders.

To account for the reduction in both circuit depth and width, we introduce a \textbf{Figure of Merit (FOM)}: defined as the product of circuit width (also known as qubit count) and its Toffoli depth. Here, we use the Toffoli gate's depth due to its higher potential for introducing error than the CNOT gate. Lower FOM is better, indicating the usage of lower quantum resources. We plot the FOM and NMED for the proposed adders in Figure \ref{fig:Graph}. Table \ref{table:comparison} shows the drop in NMED with respect to QMA1 to help understand the gradual reduction for the proposed adders.

Most of the reduction in NMED, roughly 18.63\%, is attributed to the about 30\% reduction in the FOM as we transition from QMA1 to QMA2, highlighting the pronounced impact of the decrease in the Toffoli gate count on error reduction. As we further reduce the FOM from QMA2 to QMA3 by 17.34\%, we observe only a modest decline of 2.53\% in NMED for QMA3. This implies that while QMA3 (12 qubits) saves 25\% qubits compared to QMA2 (16 qubits), its effect on error resilience is less significant than gate depth optimization. We calculate all reductions in NMED and FOM relative to QMA1, the first proposed design. Significantly, we witness another 7.64\% drop in NMED over existing reductions when the count of zero resets doubles from 5 in QMA3 to 10 in QMA4, showing that the extra zero resets successfully increase the error resilience of QMA4. Initialization of purer \text{$|0$}⟩ states by additional zero resets in the quantum half adder of QMA4 mitigates the propagation of error to MSB, even though both QMA3 and QMA4 have the same gate and qubit count.

\section{Conclusion}
In this work, we propose four novel designs of quantum modulo (2\textsuperscript{n} + 1) adder, two each in static and dynamic quantum circuits categories, a significant advancement in quantum modular arithmetic. The proposed adders provide both regular and Modulo (2\textsuperscript{n} + 1) Sums concurrently, addressing a crucial requirement in quantum arithmetic. Our first design, QMA1, is a novel design targeted specifically for quantum modulo (2\textsuperscript{n} + 1) addition. The second design, QMA2, optimizes QMA1 using a quantum half adder, reducing gate count and depth for both CNOT and Toffoli gates. QMA1 and QMA2 are static quantum circuits and hence fully reversible. The third proposed adder, QMA3, eliminates the need for (n+1) ancillae lines by applying zero resets on qubits carrying one of the inputs, leading to a 25\% qubit count reduction. QMA4 reduces state preparation error by utilizing twice as many zero resets as QMA3. Both QMA3 and QMA4 are dynamic quantum circuits with non-reversible zero reset operations.

We demonstrate the error resilience of all four proposed quantum modulo (2\textsuperscript{n} + 1) adders by experiments on the IBM Washington quantum computer, a 127-qubit system based on Eagle R1 architecture. We demonstrate a 28.8\% decrease in error with systematic design optimization of quantum modulo (2\textsuperscript{n} + 1) adders from QMA1 to QMA4. The error reduction is attributed to the decreased gate count and reduced quantum state preparation error. Ultimately, QMA2 and QMA4 are superior in their respective categories. We expect our work to lay the foundation for quantum circuits for subtraction, multiplication, and exponentiation within the modulo (2\textsuperscript{n} + 1) arithmetic. We conclude that as newer applications of quantum computing involving modulo arithmetic emerge, our proposed quantum modulo (2\textsuperscript{n} + 1) adders will find applications in diverse fields, including residue number systems based applications, quantum signal/image processing, and quantum cryptography.
 
\vspace{-0.4cm}
\bibliographystyle{IEEEtran}
\bibliography{IEEEabrv, references}

\begin{thebibliography}{10}
\providecommand{\url}[1]{#1}
\csname url@samestyle\endcsname
\providecommand{\newblock}{\relax}
\providecommand{\bibinfo}[2]{#2}
\providecommand{\BIBentrySTDinterwordspacing}{\spaceskip=0pt\relax}
\providecommand{\BIBentryALTinterwordstretchfactor}{4}
\providecommand{\BIBentryALTinterwordspacing}{\spaceskip=\fontdimen2\font plus
\BIBentryALTinterwordstretchfactor\fontdimen3\font minus
  \fontdimen4\font\relax}
\providecommand{\BIBforeignlanguage}[2]{{%
\expandafter\ifx\csname l@#1\endcsname\relax
\typeout{** WARNING: IEEEtran.bst: No hyphenation pattern has been}%
\typeout{** loaded for the language `#1'. Using the pattern for}%
\typeout{** the default language instead.}%
\else
\language=\csname l@#1\endcsname
\fi
#2}}
\providecommand{\BIBdecl}{\relax}
\BIBdecl

\bibitem{jiang2014quantum}
N.~Jiang, W.-Y. Wu, and L.~Wang, ``The quantum realization of arnold and
  fibonacci image scrambling,'' \emph{Quantum information processing}, vol.~13,
  no.~5, pp. 1223--1236, 2014.

\bibitem{ceschini2023modular}
A.~Ceschini, A.~Rosato, and M.~Panella, ``Modular quantum circuits for secure
  communication,'' \emph{IET Quantum Communication}, vol.~4, no.~4, pp.
  208--217, 2023.

\bibitem{roetteler2017quantum}
M.~Roetteler, M.~Naehrig, K.~M. Svore, and K.~Lauter, ``Quantum resource
  estimates for computing elliptic curve discrete logarithms,'' in
  \emph{Advances in Cryptology--ASIACRYPT 2017}.

\bibitem{putranto2022another}
D.~S.~C. Putranto, R.~W. Wardhani, H.~T. Larasati, and H.~Kim, ``Another
  concrete quantum cryptanalysis of binary elliptic curves,'' \emph{Cryptology
  ePrint Archive}, 2022.

\bibitem{cuccaro2004new}
S.~A. Cuccaro, T.~G. Draper, S.~A. Kutin, and D.~P. Moulton, ``A new quantum
  ripple-carry addition circuit,'' \emph{arXiv preprint quant-ph/0410184},
  2004.

\bibitem{draper2004logarithmic}
T.~G. Draper, S.~A. Kutin, E.~M. Rains, and K.~M. Svore, ``A logarithmic-depth
  quantum carry-lookahead adder,'' \emph{arXiv preprint quant-ph/0406142},
  2004.

\bibitem{rodney2005}
R.~Van~Meter and K.~M. Itoh, ``Fast quantum modular exponentiation,''
  \emph{Phys. Rev. A}, vol.~71, p. 052320, May 2005.

\bibitem{thapliyal2013design}
H.~Thapliyal and N.~Ranganathan, ``Design of efficient reversible logic-based
  binary and bcd adder circuits,'' \emph{ACM Journal on Emerging Technologies
  in Computing Systems (JETC)}, vol.~9, no.~3.

\bibitem{mohan2016residue}
P.~A. Mohan, \emph{Residue Number Systems}.\hskip 1em plus 0.5em minus
  0.4em\relax Springer, 2016.

\bibitem{nielsen2002quantum}
M.~A. Nielsen and I.~Chuang, \emph{Quantum computation and quantum
  information}.\hskip 1em plus 0.5em minus 0.4em\relax American Association of
  Physics Teachers, 2002.

\bibitem{liang2012new}
J.~Liang, J.~Han, and F.~Lombardi, ``New metrics for the reliability of
  approximate and probabilistic adders,'' \emph{IEEE Transactions on
  computers}, vol.~62, no.~9, pp. 1760--1771, 2012.

\bibitem{bgaur2023noise}
B.~Gaur, T.~Humble, and H.~Thapliyal, ``Noise-resilient and reduced depth
  approximate adders for nisq quantum computing,'' in \emph{Proceedings of the
  Great Lakes Symposium on VLSI 2023}, 2023, pp. 427--431.

\bibitem{Qiskit}
{Qiskit contributors}, ``Qiskit: An open-source framework for quantum
  computing,'' 2023.

\end{thebibliography}


\end{document}